\documentclass[12pt,twoside,draftclsnofoot,onecolumn]{IEEEtran}

\ifCLASSINFOpdf
\else
\fi

\usepackage{graphicx,subfigure,array,amssymb,latexsym,amssymb,lipsum,algpseudocode,epsfig,float,epsf,amsmath,epstopdf,cite,tikz, float,algpascal}

\usepackage[linesnumbered,ruled,vlined]{algorithm2e}% http://ctan.org/pkg/algorithm2e
\DontPrintSemicolon

\DeclareGraphicsExtensions{.pdf,.png,.jpg,.eps}

\makeatletter
\newcommand{\removelatexerror}{\let\@latex@error\@gobble}
\makeatother

\makeatletter
\def\BState{\State\hskip-\ALG@thistlm}
\makeatother
\begin{document}
%\title{Optimal CCA Threshold for Maximal Throughput in Dense IEEE 802.11 Networks}
%\title{AP Selection Algorithm for Downlink Throughput Enhancement in Dense IEEE 802.11 Networks}

\title{Decentralized AP Selection in Large-Scale Wireless LANs Considering Multi-AP Interference}

%\title{Inter-AP Interference Aware AP Selection Scheme in Large-Scale Wireless Local Area Networks}

%\title{New Channel Sounding AP Selection Scheme for Throughput Maximization in Dense Wireless LANs}

%\title{AP Selection for Downlink Throughput Enhancement in Dense Wireless LANs}

%\title{Optimal CCA Threshold for Performance Enhancement in Dense IEEE 802.11 Networks}
%\title{CCA Threshold Optimization for Performance Enhancement in Dense WLANs}

%Enhancing Uplink Throughput in Dense WLANs via Association and CCA threshold Optimization

%Association and CCA Threshold Optimization to Enhance Uplink Throughputs in Dense WLANs
%\author{\IEEEauthorblockN{Phillip B. Oni, \textit{Student Member, IEEE} and Steven D. Blostein, \textit{Senior Member, IEEE}}
%\IEEEauthorblockA{Department of Electrical and Computer Engineering\\
%Queen's University, Kingston, ON, Canada.\\
%Email: \{phillip.oni, steven.blostein\}@queensu.ca}}

\author{\IEEEauthorblockN{Phillip B. Oni and Steven D. Blostein}
\IEEEauthorblockA{Department of Electrical and Computer Engineering\\
Queen's University, Kingston, ON, Canada.\\
Email: \{phillip.oni, steven.blostein\}@queensu.ca}}


%Telephone: (800) 555--1212, Fax: (888) 555--1212}


\maketitle

\begin{abstract}
\noindent Densification of access points (APs) in wireless local area networks (WLANs) increases the interference and the contention domains of each AP due to multiple overlapped basic service sets (BSSs). Consequently, high interference from multiple co-channel BSS at the target AP impairs system performance. To improve system performance in the presence of multi-BSSs interference, we propose a decentralized AP selection scheme that takes interference at the candidate APs into account and selects AP that offers best signal-interference-plus noise ratio (SINR). In the proposed algorithm, the AP selection process is distributed at the user stations (STAs) and is based on the estimated SINR in the downlink. Estimating SINR in the downlink helps capture the effect of interference from neighboring BSSs or APs. Based on a simulated large-scale 802.11 network, the proposed scheme outperforms the strongest signal first (SSF) AP selection scheme used in current 802.11 standards as well as the mean probe delay (MPD) AP selection algorithm in \cite{chen}; it achieves 99\% and 43\% gains in aggregate throughput over SSF and MPD, respectively. While increasing STA densification, the proposed scheme is shown to increase aggregate network performance.\\

%Enhancing downlink (DL) throughput is crucial because it is envisaged that in dense IEEE 802.11 networks, the majority of network traffic will be in the downlink especially in the emerging dense deployment of APs for cellular-WiFi offloading. 

%there is 75\% gain in per basic service set (per-BSS) DL throughput while the peak-PHY and mean-PHY rates increase by 56\% and 80\% respectively. 

%Finally, results show that when STA-AP association is coordinated and CCA threshold is adjusted , STAs' throughputs significantly improve.
\end{abstract}


\section{Introduction}

%Listen-to-yourself (LTY) interference avoidance scheme in Dense CSMA Network
The popularity of IEEE 802.11 or Wireless Fidelity (Wi-Fi) networks among users as an affordable data access network is increasing tremendously, and consequently, causing an increase in the number of access points (APs) deployed in places like residential/apartment buildings, hotels, airports, campuses and enterprise buildings. The unprecedented demand for affordable high data rate and the emergence of bandwidth intensive applications is also a contributing factor. Similarly, the emerging cellular-WiFi offloading trend requires a high density of APs to handle the huge mobile data traffic \cite{klee}. With this promising solution to explosive mobile data traffic comes increased inter-AP interference, which degrades the capacity of dense wireless local area networks (DWLANs).

Although densification of APs provides extended coverage and affordable data access in homes, offices and campuses, deploying large numbers of APs over a confined network area increases the interference domain of each AP and causes severe interference to neighboring APs. This becomes worrisome in cases where AP cells overlap leading to overlapped basic service sets (OBSS) where inter-BSS or inter-AP interference becomes significant \cite{shin}; a BSS consists of an AP and its associated stations (STAs). In addition to increasing AP's interference domain, uncoordinated distribution of STAs among APs causes overwhelming channel access contention at overloaded cells due to the carrier sense multiple access collision avoidance (CSMA/CA) protocol specified in the IEEE 802.11 standard for channel acquisition. Interference, data collisions and congestion are major concerns in DWLAN. 

One technique to improve system performance in the presence of multiple interference sources is to coordinate AP selection based on key performance metrics such as SINR and packet error rate (PER). Currently, the AP selection process in WLAN is based on the strongest received signal strength (RSS) or strongest signal first (SSF), a method whereby an STA selects AP that offers strongest RSS without considering interference, congestion and load at the candidate AP. This legacy AP selection scheme defined in the 802.11 standard might cause a high degree of contention in some BSSs, and consequently degrades aggregate network performance. Studies \cite{chen} - \cite{bjovic} focus on proposing new schemes for AP selection in 802.11 networks and demonstrate the inability of an RSS-based scheme to guarantee a level system performance.

\section{Existing AP Selection Schemes}

In \cite{chen} and \cite{weili}, authors focus primarily on achieving a fair distribution of STAs among APs to achieve load balancing as opposed to the SSF AP selection scheme that has the tendency to cause load imbalances among APs \cite{hong}. The probe delay (PD) and mean probe delay (MPD) algorithms\cite{chen} select the AP with minimum probe delay. Similarly, an AP association control scheme is proposed in \cite{weili} to achieve proportional fairness. A graph matching approach to coordinate AP association and maximize uplink throughput is proposed in \cite{pbo}, where the links between STAs and APs are modeled as graph edges with uplink SINRs as edge weights. In \cite{masahiro}, virtualizing wireless network interfaces enables STAs to associate with multiple APs and switch between APs without severe overhead thereby making the AP selection dynamic.

In \cite{leidu}, by introducing a Quality of Service (QoS) differentiated information element (IE) in frames advertised by APs, STAs are aware of the \textit{call blocking} probability when selecting an AP. By measuring channel utilization, the authors in \cite{leidu2} proposed an AP selection scheme that allows STAs to select AP with minimum hidden terminal effect. Similarly, using a channel measurement approach, the authors in \cite{hong} suggest that the hidden terminal problem and frame aggregation are factors in selecting an AP that guarantees better throughput. Another measurement-based AP selection approach is presented in \cite{bjovic} using a supervised learning technique (Multi-Layer Feed-Forward Neural Network) with multiple inputs, which allows STAs to select the AP that offers best performance.

\section{Contribution}

Thus far, inter-BSS interference at the target AP has not been considered when selecting AP. The proposed AP selection scheme exploits awareness of inter-AP interference by allowing STAs to estimate the received SINRs from a set of candidate APs and select an AP with the best SINR. The goal is to improve system performance by associating STAs with APs that offer best SINR. Enhancing performance in the presence of inter-AP interference is important in DWLANs for two reasons. First, the problem of OBSS is inevitable in large-scale dense AP deployments, leading to severe inter-AP interference, which degrades performance due to close proximity of co-channel APs. 

Second, the majority of the traffic is in the downlink. Taking video streaming as an example, after a user requests the service in the UL, the entire video streaming session occurs in the downlink.  The remaining parts of this paper are organized as follows. First, we present the system and network model in Section~\ref{sysmodel}. The proposed scheme is presented in Sections~\ref{dlassoc} and \ref{opasa} while simulation results are presented in Section~\ref{eval}. Section~\ref{conclusion} concludes this paper. The summary of key symbol definitions is presented in Table~\ref{table:param} for easy reference.

\begin{table}[!h]
     \centering
     \caption{Key Notations.}
     %\scriptsize
     %\resizebox{\columnwidth}{!}
    {\begin{tabular}{|l|l| }
          \hline
          \textbf{Notation} & \textbf{Definition} \\ 
      \hline
      	${\mathcal{S}}$ & Set of stations (STAs) \\
      \hline
      	${\mathcal{A}}$ & Set of access points (APs)\\
      \hline
      $M = |{\mathcal{A}}|$ & Total number of APs \\
      \hline
      $N = |{\mathcal{S}}|$ & Total number of STAs \\ 
     \hline
     $\Gamma$ & Physical carrier sensing (PCS) threshold \\
	\hline
	$\omega$ & A channel from set of orthogonal channels \\
	\hline
	$P^t$ & Transmit power \\
	\hline
	$\mathcal{A}^{\omega}$ & Set of co-channel APs on channel $\omega$\\
	\hline
	${\mathcal{A}}^{\omega}_{a}$ & Set of active APs (permitted by CSMA/CA to transmit)\\
	\hline
	${\mathcal{A}}^I$ & Set of APs interfering \\
	\hline
	${\mathcal{A}}_{i}^c$ & Set  of candidate APs \\
	\hline
	$P^r_{ji}$ & Received power by STA$_{i}$ from AP$_{j}$ \\
     \hline
     	$\gamma_{o}$ & SINR threshold\\
     	\hline
	$\theta$ & Receiver sensitivity \\
     	\hline
     	${\mathcal{I}}_{T}^{ji}$ & Total interference power on a desired link\\
     	\hline
     	$t_{ji}$ & Transmission time of a frame \\
     	\hline
     	$\Lambda_{ji}$ & Transmission rate \\
     	\hline
     	${\mathbf{B}}_{ji}$ & Expected throughput of a link \\
     	\hline
    \end{tabular}
     \label{table:param}}
    \end{table}

\section{System Model and Problem Formulation}{\label{sysmodel}}

\begin{figure}[!h]
	\centering
	\includegraphics[width=5in]{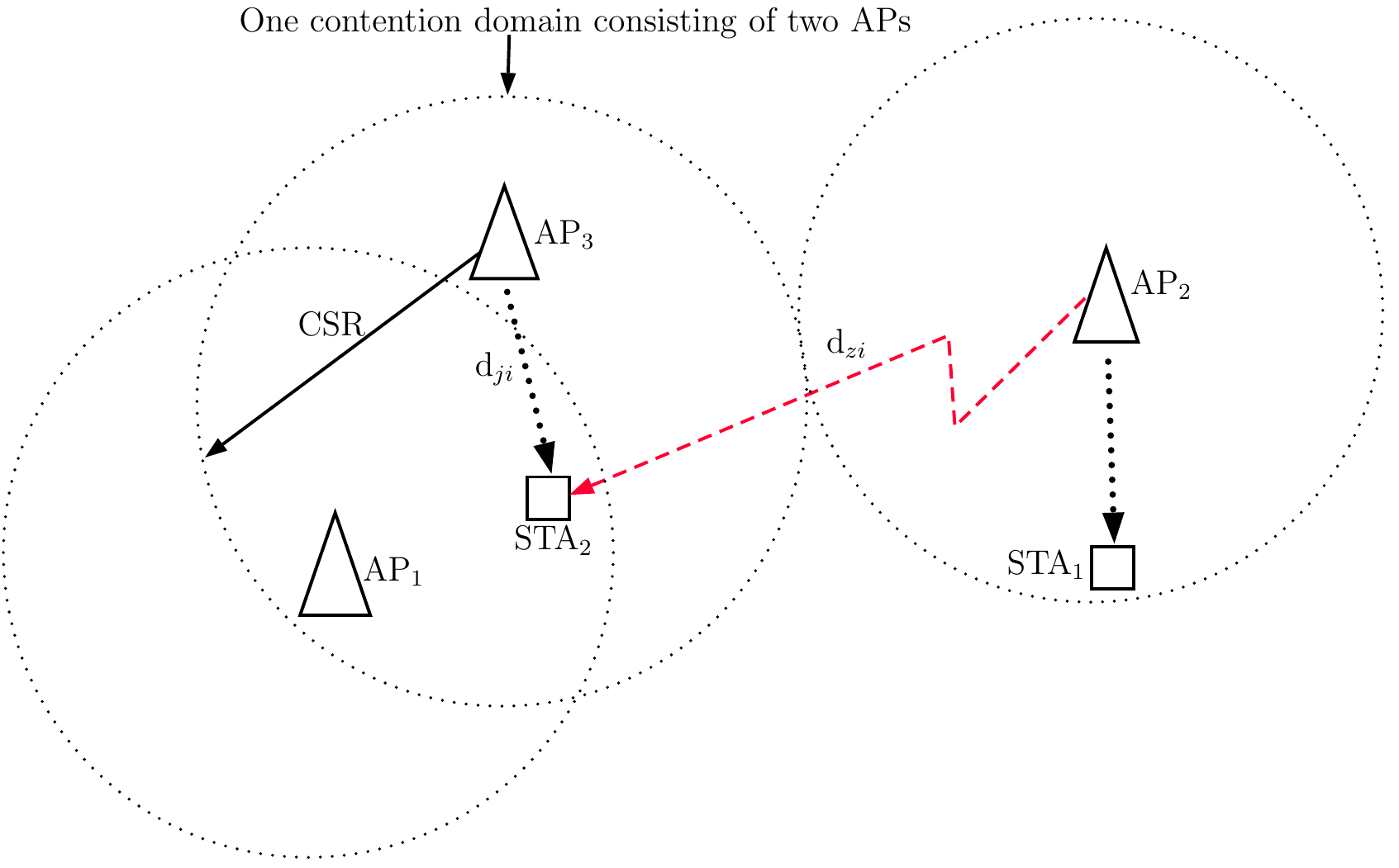}
	\caption{System and interference model.}
	\label{fig:dl}
\end{figure}

In this section, the system model for a DL (AP-to-STA) transmission is presented. In the downlink (DL) of a WLAN, APs transmit data to their respective associated STAs as shown in our system model in Figure~\ref{fig:dl}. The achievable throughput in the downlink is of particular concern because the majority of dense Wi-Fi traffic will be in the downlink. On a typical DWLAN, let the set of APs be denoted as ${\mathcal{A}}$ serving a set ${\mathcal{S}}$ of STAs. Hence, the entire network consists of $M = |{\mathcal{A}}|$ APs and $N = |{\mathcal{S}}|$ STAs. For this downlink model, we will assume that all APs transmit at the same power $P^t \left(\mbox{mW}\right)$ and $d_{ji}$ ($\forall j\in {\mathcal{A}}, i \in {\mathcal{S}}$) is the distance between the transmitting AP$_j$ and the receiving STA$_i$ as shown in Figure~\ref{fig:dl}. 

Next, we describe channel contention in the downlink of a WLAN. The number of available orthogonal channels in Wi-Fi networks depends on the version of the IEEE 802.11 standard being supported. For instance, the IEEE 802.11b/g standard supports 3 non-overlapping channels from 14 available channels while IEEE 802.11a provides 8 non-overlapping channels. The more recent standard, IEEE 802.11ac, has two non-overlapping channels for 80MHz and one 160MHz non-overlapping channel. Therefore, due to an insufficient number of orthogonal channels for large AP deployments, many APs are deployed on the same channel and some BSSs overlap due to close proximity. Consequently, two or more APs must contend for the same channel before transmitting.

In Figure~\ref{fig:dl}, let AP$_3$ and AP$_1$ represent co-channel APs within carrier sensing range (CSR) of each other. The CSR depends on the clear channel assessment (CCA) threshold used during the physical carrier sensing (PCS) process. PCS is usually performed within the CSR to determine the presence of active transmissions on the channel. In order for an AP to detect the presence of an active AP on the channel, the energy level sensed during PCS is compared to the CCA threshold. The channel is occupied by another AP within the CSR if the sensed energy level is greater than the CCA threshold. Therefore, with the PCS process in CSMA/CA, whenever AP$_3$ has the channel for transmission, AP$_1$ remains silent; all co-channel APs do not transmit concurrently.

For any supported 802.11 standard, let $\omega$ denote a channel belonging to the set of available orthogonal channels. Let us denote the set of co-channel APs on channel $\omega$ as $\mathcal{A}^{\omega}$ and let ${\mathcal{A}}^{\omega}_{a}$ represent the set of active APs (permitted by CSMA/CA to transmit) in ${\mathcal{A}}^{\omega}$ on channel $\omega$. By virtue of the CCA threshold, a subset of the active APs in ${\mathcal{A}}^{\omega}_{a}$ will be in the contention domain (within CSR) of AP$_{j}$, $\forall j \in {\mathcal{A}}$. Therefore, all active APs in the contention domain of AP$_{j}$ form a set of co-channel APs with AP$_{j}$ and this set is denoted as ${\mathcal{A}}^{\omega}_{j}$. This implies that AP$_{j}$ will contend for the medium with other co-channel APs in ${\mathcal{A}}^{\omega}_{j}$, and if AP$_{j}$ or any other AP in ${\mathcal{A}}^{\omega}_{j}$ is transmitting on the channel, other APs are idle; mathematically: ${\mathcal{A}}^{\omega}_{j} := \{ m \in {{\mathcal{A}}^{\omega}_{a}}, m \neq j | \lambda > \Gamma, {\mathcal{A}^{\omega}}  \subseteq {\mathcal{A}}\}, j \in {\mathcal{A}}, \omega \in {\mathcal{C}}$, where $\Gamma$ denotes the PCS (or CCA) threshold, $\lambda$ is the signal power sensed on channel $\omega$ during the PCS, and ${\mathcal{C}}$ is the set of channels in any supported IEEE 802.11 standard. All APs within the CSR of AP$_j$ are not potential interference sources because the CSR area is cleared during PCS and all other APs within the CSR do not transmit while AP$_j$ is active.

%\begin{equation}
%%\small
%\begin{split}
%\label{intset}
%{\mathcal{A}}^{\omega}_{j} := \{ m \in {{\mathcal{A}}^{\omega}_{a}}, m \neq j | \lambda > \Gamma, {\mathcal{A}^{\omega}}  \subseteq {\mathcal{A}}\}, j \in {\mathcal{A}}, \omega \in {\mathcal{C}}
%\end{split}
%\end{equation}

Figure~\ref{fig:dl} illustrates a downlink interference scenario where AP$_2$ is outside the contention domain of AP$_{3}$. Therefore, a signal coming from AP$_{2}$ might interfere with downlink transmissions of AP$_{3}$ at receiver STA$_{2}$. From this scenario, the total interference at the receiving STA$_{2}$ in the downlink can be estimated depending on the number of APs transmitting outside the CSR of AP$_{3}$ and whose signals are received by STA$_{2}$. Let ${\mathcal{A}}^I$ represent the set of APs interfering with the downlink signal of AP$_{j}$ at receiver STA$_{i}$. The total interference received at STA$_{i}$ from all interfering APs in ${\mathcal{A}}^I$ is given by
\begin{equation}
%\small
%\scriptsize
{\mathcal{I}}^{ji} = \sum_{z \in {\mathcal{A}}^I, i \in {\mathcal{S}}, j \in {\mathcal{A}} | j \neq z, {\mathcal{A}}^I \subset {\mathcal{A}}^{\omega}_{j} \subset {\mathcal{A}}}^{} P_{zi},
\label{dlint}
\end{equation}

\noindent where $P_{zi}$ is the received signal power at STA$_{i}$ from the $z^{th}$ interfering AP at distance $d_{zi}$. This type of interference measurement is based on a one time capture of the signal strength of an interference source. It does not account for the time variations of the wireless channel and signal strength. Also, the frames received from different interfering sources vary in size and each interfering AP might use a different PHY rate for transmission. Therefore, using the \textit{passive interference measurement} approach in \cite{murad}, we can reformulate (\ref{dlint}) to account for these variations as follows:
\begin{equation}
%\small
\label{dlint2}
{\mathcal{I}}_{T}^{ji} = \frac{1}{T} \sum_{z = 1}^{|{\mathcal{A}}^I|} \sum_{k = 1}^{K}\frac{P_{zi} L_{zi}}{R_{zi}}, \mbox{ } z \in {\mathcal{A}}^I, i \in {\mathcal{S}}, j \in {\mathcal{A}} | j \neq z
\end{equation}

\noindent where $K$ is the number of frames and $L_{zi}$ is the length of each frame in bits received from $z^{th}$ interferer, $R_{zi}$ is the PHY rate (bps) at which each frame is received and $T$ denotes the measurement period, and can represent a sufficient number of slot times for accurate estimation. As a result of interfering signal power received at STA$_{i}$ from APs outside AP$_{j}$'s CSR, the SINR of the link between AP$_{j}$ and STA$_{i}$ is given by
\begin{equation}
%\small
%\scriptsize
\Psi_{ji} = \frac{P^r_{ji}}{\left({\mathcal{I}}_{T}^{ji} + N_{o}\right) W}, \qquad 1 \leq i \leq N, 1 \leq j \leq M,
\label{snr}
\end{equation}

\noindent where $P^r_{ji}$ is the received power from AP$_{j}$ at STA$_{i}$ over a distance $d_{ji}$ and $W$ is the system bandwidth. The total transmission time of a desired frame of size $\digamma$ from AP$_{j}$ to STA$_{i}$ is denoted as
\begin{equation}
t_{ji} = \frac{\digamma \left(\mbox{bits}\right)}{\Lambda_{ji}},
\end{equation}

\noindent where $\Lambda_{ji}$ is the transmission rate, which is determined by SINR $\Psi_{ji}$ experienced by STA$_{i}$ when associated with AP$_{j}$. The mapping or relationship between $\Lambda_{ji}$ and $\Psi_{ji}$ in an 802.11 WLAN is shown in Table~\ref{table:tableII}. Denote the expected throughput of STA$_i$ after associating with AP$_{j}$ as:

\begin{equation}
{\mathbf{B}}_{ji} =  \frac{1}{t_{ji}},
\end{equation}

\noindent which is the case provided prolonged transmission time due to retransmission (aftermath of collision) does not occur. However, when a frame from AP$_{j}$ to STA$_i$ experiences collision, the transmission time is prolonged as thus:
\begin{equation}
\tilde{t}_{ji} = \mbox{DIFS} + t_{bf} + t_{ji} + \mbox{SIFS} + t_{ack},
\end{equation}

\noindent where $t_{bf} = \frac{\mbox{CW}_{\max}}{2} \times \mbox{Slot-time}$ is the backoff time, $t_{ack} =  \frac{1}{r}$ is the time it takes to transmit ACK frame given basic data rate $r$ (e.g. 1Mbps in an 802.11b network) while $\mbox{SIFS}$ and $\mbox{DIFS}$ are time intervals defined in the 802.11 standard.
 
%Using Shannon's capacity, the DL PHY rate is upper bounded by:
%\begin{equation}\label{dlthrput}
%%\small
%%\scriptsize
%\Lambda_{ji} = W \log\left(1 + \Psi_{ji}\right).
%\end{equation}

  \begin{table}[!h]
     \centering
     \caption{SINR requirements for different data rates in 802.11 \cite{weili}.}
     \scriptsize
     \resizebox{\columnwidth}{!}
    {\begin{tabular}{|c|c|c| c |c|c|c| c | c | }
          \hline
          $\Psi_{ji}$(dB) & 6-7.8 & 7.8-9 & 9-10.8 & 10.8-17 & 17-18.8 & 18.8-24 & 24-24.6 & $>24.6$\\ 
      \hline
      $\Lambda_{ji}$(Mbps) & 6 & 9 & 12 & 18 & 24 & 36 & 48 & 54\\  
      %\hline
      %\textbf{Minimum sensitivity, $\theta$} (dBm) & -65 & -66 & -70 & -74 & -77 & -79 \\ 
      \hline
    \end{tabular}
     \label{table:tableII}}
    \end{table}
 
\section{AP Selection Algorithm}\label{dlassoc}

In this section, we present the proposed AP selection method as \textbf{Algorithm~\ref{dlassocAl}}.  The \textit{probe request} and \textit{probe response} frames defined in IEEE 802.11 standards are used to perform interference measurement in the downlink. A typical STA$_i$ captures the beacon frames (through \textit{channel scanning}) from all APs within range to determine the set  of candidate APs, ${\mathcal{A}}_{i}^c$, and selects the best-serving AP. Let $\kappa$ be the set of APs within range of STA$_{i}$. In Step 2, a typical STA$_i$ listens to beacon frames from all APs within range and sorts the RSSs of the beacon frames in decreasing order (Step 3) and selects the AP with best RSS to complete SSF association.

The SSF association is important to ensure that all STAs can discover APs within range and continue to transmit/receive payloads while estimating SINRs from all candidate APs. This prevents starvation in cases when a STA cannot find the AP offering best SINR. Once SSF association is achieved by STA$_i$, it proceeds in Step 4 to create set ${\mathcal{A}}_{i}^c$ of candidate APs for STA$_{i}$ from the set of APs within range. An AP is added to ${\mathcal{A}}_{i}^c$ if its RSS at STA$_{i}$ satisfies the minimum receiver sensitivity constraint i.e., $ P^r_{ji} > \theta$. The choice of $\theta$ depends on the supported data rate. For example, in a typical 802.11 network, to support a minimum data rate of 12 Mbps, the receiver sensitivity $\theta$ is -79 dBm (\textit{for successful reception}). After obtaining the set of candidate APs, the channel measurement using probe request and response frames begins in Step 5.

%\vspace{-.3cm}
\begin{figure}[!h]
\removelatexerror
\begin{algorithm}[H]
%\scriptsize
  \KwIn{$\theta$, ${\mathcal{S}}$, ${\mathcal{A}}$}
  \KwOut{$ (x_{ji})_{i \in {\mathcal{S}}, j \in {\mathcal{A}}} \rightarrow \textbf{x} $} 
  \textbf{Initialization:} STAs associate with AP offering best RSS\;
  Typical STA$_{i},$ $\forall i \in {\mathcal{S}}$ listens to APs' beacon frames\;
  STA$_{i}$ captures RSS from APs within range: $P^r_{1i} \geq P^r_{2i} \geq \dots P^r_{|\kappa|i}$ \;
  \For{$j \leq |\kappa|$ (For each AP within range); $\forall j \in \kappa, \kappa \subseteq {\mathcal{A}}$} {\textbf{if} $P^r_{ji}$ $\geq \theta$; $\forall j \in \kappa, \kappa \subseteq {\mathcal{A}}, \forall i \in {\mathcal{S}} $ \; $\qquad$ STA$_i$ adds AP$_j$ to set ${\mathcal{A}}_{i}^c$ of candidate APs\; \textbf{end if}}
  \For{$j \leq |{\mathcal{A}}_{i}^c|$ (For each candidate AP for STA$_i$)}{STA$_i$ sends \textit{probe request frame} (\texttt{P\_REQ}) to AP$_j$ \; \While{$T \leq n \times \mbox{Slot time}$}{AP$_j$ sends \texttt{P\_RES} to STA$_i$ \; STA$_i$ captures the power level ($\mbox{dBm}$) from AP$_j$ \; STA$_i$ estimates the interference power ($\mbox{dBm}$) arriving with \texttt{P\_RES} \; } Using (\ref{dlint2}), STA$_i$ estimates the average interference power level ($\mbox{dBm}$) \; STA$_i$ estimates DL-SINR from AP$_j$, $\Psi_{ji}$ using (\ref{snr}) \; STA$_{i}$ sends {\texttt{ASSOC$\_$REQ}} frame to AP$_{j}$ offering best DL-SINR: $\hspace{\algorithmicindent}\Psi_{1i} \geq \Psi_{2i} \geq \dots \Psi_{\kappa N }$ }
  AP$_j$ replies with {\texttt{ASSOC$\_$RES}} frame \; STA$_{i}$ associates or re-associates with AP$_j$
 \caption{DL-SINR AP Selection Algorithm (DASA)}
  \label{dlassocAl}
\end{algorithm}
\end{figure}
%\vspace{-.3cm}

The typical STA$_{i}$ sends a directed \textit{probe request frame} to each AP in ${\mathcal{A}}_{i}^c$. To increase the accuracy of the interference measurement in Step 5, the algorithm requires that multiple \texttt{P\_RES} frames be received by STA$_i$ over a specified period of time. To do this, we define a measurement period $n \times \mbox{Slot-time}$ where $n$ is an integer. Due to the power constraint of most STAs and other low power devices e.g., WiFi-enabled Internet of things (IoTs), \textbf{Algorithm~\ref{dlassocAl}} requires that STAs send only one \texttt{P\_REQ} frame but informs candidate AP$_j$ to send multiple \texttt{P\_RES} frames within $n \times \mbox{Slot-time}$. The \textit{RequestInformation} (\texttt{dot11RadioMeasurementActivated} = true) parameter of the MLME-Scan.request primitive defined for channel scanning in IEEE 802.11 standard (2012)  can be used to inform a candidate AP$_j$ to send multiple \texttt{P\_RES} frames.

On receiving the \textit{probe response frame} (\texttt{P\_RES}) from AP$_j$, STA$_{i}$ estimates the DL-SINR based on the magnitude of captured interference power received concurrently with \texttt{P\_RES}. Alternatively, this SINR estimation can be done through \textit{channel sounding} by sending preamble frames or training symbols to candidate APs. Then, it sends association request (\texttt{ASSOC\_REQ}) frame to the candidate AP that offers the best DL-SINR. The candidate AP responds with association response (\texttt{ASSOC\_RES}) frame. This algorithm is easy to implement and does not require modification to 802.11 management frames. Also, we remark that channel measurement capability is available in 802.11k-enabled nodes. 

\section{Optimal AP Selection Algorithm}{\label{opasa}}

The AP selection scheme in Section~\ref{dlassoc} might not be optimal under SINR, receiver sensitivity and CCA threshold constraints. These constraints are related to interference distribution across the network. In this section, the problem of AP selection to maximize aggregate throughput is formulated as follows:
\begin{subequations}
%\small
\label{dlassoc_eq}
\begin{align}
       	  	  	& \text{maximize} & &  \sum_{j = 1}^{M}\sum_{i=1}^{S}{\mathbf{B}}_{ji}x_{ij} \label{dla} \\
       	  	  	& \text{subject to} & & \sum_{j = 1}^{M} x_{ij} = 1, \forall i \in {\mathcal{S}} \label{dlb}\\
       	  	  	& & & x_{ij}P^r_{ji} \geq \theta \qquad \forall j \in {\mathcal{M}} \label{rss}\\
       	  	  	%& & & \sum_{j = 1}^{M} {\mathcal{I}}^{ij}x_{ij} \leq \Gamma, \forall i \in {\mathcal{N}}\\
       	  	  	& & & x_{ij}\Psi_{ji} \geq \gamma_{o} \qquad \forall i \in {\mathcal{S}} \label{dlc}\\
       	  	  	& & & x_{ij}P^c_{ji}  \leq \Gamma \qquad \forall i \in {\mathcal{S}} \label{dld},
\end{align}
\end{subequations}

\noindent where $x_{ij} \in \{0,1\}, i \in {\mathcal{S}}, j \in {\mathcal{A}}$, $P^c_{ji}$ is the total power sensed on the channel during PCS, $\gamma_{o}$ is the SINR threshold, constraint (\ref{dlb}) ensures that each STA associates with only one AP; $x_{ij} = 1$ if STA$_{i}$ is associated with AP$_{j}$ and $x_{ij} = 0$ if otherwise. (\ref{rss}) is the receiver sensitivity constraint while (\ref{dlc}) and (\ref{dld}) are the SINR and CCA threshold constraints,  respectively. An AP begins transmission if (\ref{dld}) is satisfied during carrier sensing. This occurs when the total interference power $P^c_{ji}$ received from other interfering APs does not exceed $\Gamma$. Therefore, $P^c_{ji} = {\mathcal{I}}^{ji}$ and SINR $\Psi_{ji}$ is related to ${\mathcal{I}}^{ji}$ by (\ref{snr}). Therefore, any feasible $\left(\Psi_{ji}\right)_{i \in {\mathcal{N}}, j \in {\mathcal{A}}}$ that satisfies (\ref{dlc}) also satisfies (\ref{dld}), hence, constraint (\ref{dld}) becomes redundant. Similarly, since $P^r_{ji}$ and $\Psi_{ji}$ are also related by (\ref{snr}) and STA$_i$ selects AP$_{j}$ offering best SINR, i.e, 
\begin{equation}
%\small
\begin{split}
j^{'} & = \underset{j}{\arg\min}\mbox{ }{\mathcal{I}}_{T}^{ji} \qquad   \forall j^{'} \in {{\mathcal{A}}^I}, j \in {{\mathcal{A}}},  i \in {{\mathcal{S}}} \\
%j^{*} & = \underset{j}{\arg\max}\mbox{ }P^r_{ji} \qquad   j \in {{\mathcal{A}}},  i \in {{\mathcal{N}}} \\
j^{*} & = \underset{j}{\arg\max}\mbox{ }\Psi_{ji} \qquad   j \in {{\mathcal{A}}},  i \in {{\mathcal{S}}},
\end{split}
\label{cond}
\end{equation}

\noindent consequently, any feasible $\left(\Psi_{ji}\right)_{i \in {\mathcal{S}}, j \in {\mathcal{A}}}$ that satisfies (\ref{dlc}) also satisfies (\ref{rss}) and renders (\ref{rss}) redundant as well. Therefore, (\ref{dlassoc_eq}) can be equivalently expressed as:
\begin{subequations}
%\small
\label{dlassoc_eq2}
\begin{align}
       	  	  	& \text{maximize} & & \sum_{j = 1}^{M}\sum_{i=1}^{S} {\mathbf{B}}_{ji}x_{ij} \label{dla2} \\
       	  	  	& \text{subject to} & & \sum_{j = 1}^{M} x_{ij} = 1, \forall i \in {\mathcal{S}},  x_{ij} \in \{0,1\} \label{dlb2}\\
       	  	  %	& & & P^r_{ji} \geq \theta \qquad \forall j \in {\mathcal{M}} \label{rss2}\\
       	  	  	%& & & \sum_{j = 1}^{M} {\mathcal{I}}^{ij}x_{ij} \leq \Gamma, \forall i \in {\mathcal{N}}\\
       	  	  	& & & x_{ij}\Psi_{ji} \geq \gamma_{o},  \forall i \in {\mathcal{S}}, j \in {\mathcal{A}}  \label{dlc2}
       	  	  %	& & & P^c_{ji}  \leq \Gamma \qquad \forall i \in {\mathcal{N}} \label{dld2}\\
       	  	   %	& & & x_{ij} \in \{0,1\}, i \in {\mathcal{S}}, j \in {\mathcal{A}}.
%& & &\Gamma_{\mathbf{a}} \leq \Gamma .
\end{align}
\end{subequations}

The solution to (\ref{dlassoc_eq2}) is given in \textbf{Algorithm~\ref{opAPselection}}, where the problem is solved numerically using linear programming (LP). A typical STA$_i$ captures SINR from all APs within range, then sets $\boldsymbol{\hat{\Psi}}$ containing SINR through each candidate AP and uses the Gurobi LP solver \cite{gurobi} to locally determine its association, $x_{ij} = 1$. The main goal of OPASA in \textbf{Algorithm~\ref{opAPselection}} is to serve as the optimal throughput benchmark given SINR constraints.
%\vspace{-.3cm}

\begin{figure}[!h]
\removelatexerror
\begin{algorithm}[H]
%\scriptsize
$\textit{initial state:} \mbox{ SSF}  $ (Steps 1 - 6 of \textbf{Algorithm~\ref{dlassocAl}})\;
STA$_i$ sets $\boldsymbol{\hat{\Psi}} = \left(\Psi_{ji}\right)_{j \in {\mathcal{A}}}, \textbf{x}$\;
 $\textbf{inputs: } \gets \boldsymbol{\hat{\Psi}}$, $\gamma_{o} = \min\left(\boldsymbol{\hat{\Psi}}\right)$\;
 $\qquad$ STA$_i$ \textbf{loads} Gurobi LP Solver: $\gets$ $\boldsymbol{\hat{\Psi}}$\;
 $\qquad$ STA$_i$ \textbf{solves} problem (\ref{dlassoc_eq2}) locally\;
 \Return $\textbf{x}$, $(\Lambda_{ji})_{i \in {\mathcal{N}}, j \in {\mathcal{A}}}$\;
 STA$_i$ $\mbox{selects}$ AP$_j$ $\iff$ $x_{ij} \in \textbf{x} = 1$
 \caption{Optimal AP Selection Algorithm (OPASA)}
   \label{opAPselection}
\end{algorithm}
\end{figure}

\begin{table*}[!t]
	\centering
	\caption{Simulation parameters}
	\label{table:simpara}
	%\scriptsize
	\begin{tabular}{ ll|ll|ll}
		%\hline
		% \multicolumn{2}{|c|}{\textbf{Symbols and Semantics}} \\
		%\hline
		\textbf{Parameter} & \textbf{Value} & \textbf{Parameter} & \textbf{Value} & \textbf{Parameter} & \textbf{Value}\\
		\hline
		Simulation network area &  $1000 \times 1000 \mbox{m}^2$   & CCA threshold, $\Gamma$ & $-86\mbox{ dBm}$ & PCS and RTS/CTS & Enabled \\
		Slot-time  & $20 \mu s$ & STA Transmit power & 15.85mW & AP Transmit Power & 100 mW\\
		$N_o$ & $90 \mbox{ dBm}$ & AP Buffer Size  & $ 20$ Packets & \texttt{P\_REQ} and \texttt{P\_RES} Frames & 20 Bytes\\
		 CCA Time / SIFS  & $15 \mu s$ / $10 \mu s$ & Receiver sensitivity, $\theta$ &$-90.96 \mbox{ dBm}$ & Mean Packet Size &  1460 bytes\\
	%	\hline
	\end{tabular}
\end{table*}

\section{Performance Evaluation}{\label{eval}}

This section presents the simulation methodology, scenario and results. For performance benchmarking, OPASA mainly serves as an optimal benchmark while DASA (Algorithm~\ref{dlassocAl}) is compared with the SSF scheme in 802.11 standards and the state-of-the-art \textit{mean probe delay} (MPD) AP selection algorithm in  \cite{chen}.

\subsection{Simulation Setup and Parameters}{\label{simset}}

 To simulate the channel access coordination at the MAC layer in a WiFi network, we have implemented the  distribution coordination function (DCF) with a Slot-time of $20 \mu s$, short inter-frame space (SIFS) $= 10 \mu s$, DIFS = $2 \times \mbox{SIFS}$ and a CCA time of $15 \mu s$ in MATLAB. The simulated network emulates a random AP deployment where APs and STAs are deployed on an area of $1000 \times 1000  \mbox{ m}^2$.  This network consists of 400 STAs and 50 APs deployed on three non-overlapping channels of IEEE 802.11b PHY on a 2.4GHz band. All APs have identical coverage areas of $50 \mbox{m}$ radius and transmit with a uniform power of $100 \mbox{mW}$ ($20 \mbox{dBm}$). Table~\ref{table:simpara} summarizes other key parameters and the received power at STA$_{i}$ from AP$j$ is measured using $P^r_{ji} = P^t \left(\mbox{mW}\right)G_{ji} d^{-\alpha}_{ji}  \left(\mbox{dBm}\right),$ where $G_{ji}$ is the channel gain characterized by an exponential distribution i.e. $G_{ji} \sim \exp\left(P^{t}\right)$ to account for fading and shadowing effects, and $\alpha = 3$ is used as the path loss exponent. The minimum receiver sensitivity is set as $\theta = -90.96 \mbox{ dBm}$.
 
For carrier sensing, the PCS is enabled with minimum and maximum contention window (CW) sizes 32 and 1024, respectively, while  request to send (RTS)/clear to send (CTS) frames are used to minimize the effect of the \textit{hidden terminal problem} - a node does not transmit immediately after sensing the channel to be idle under PCS. Rather, it transmits the RTS frame and begins transmission of payload when the CTS frame is received. To emulate the asymmetric traffic requests in Wi-Fi networks, APs and STAs transmit packets of varying sizes (between 1400 to 1500 bytes) with a mean packet size of 1460 bytes while the MAC header, CTS/RTS frame and ACK frame sizes are 34, 14/20 and 14 bytes respectively as defined in the 802.11 standard. It is assumed that the \texttt{P\_RES} and \texttt{P\_REQ} frames have same size as the RTS. Packets arrive at each node's buffer at an exponential rate with parameter $\lambda =  1/\mbox{Slot-time}$. 

\subsection{Simulation Results and Performance Benchmarking}{\label{simres}}

\begin{figure}[!h]
	\centering
	\includegraphics[width=5in]{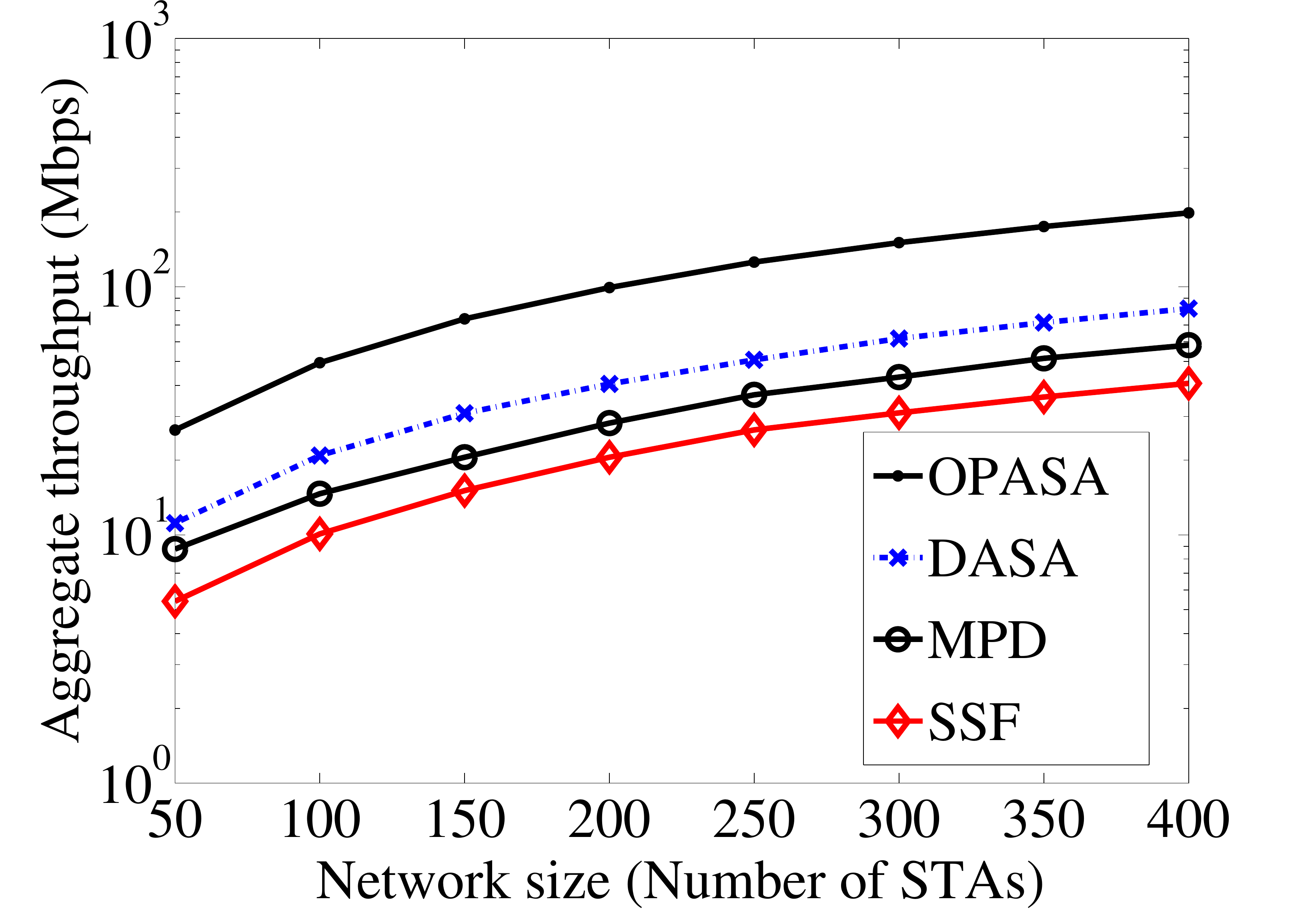}
	\caption{Aggregate end-to-end throughput versus network size for $n=1000 \mbox{ Slot times}$.}
	\label{fig:res1}
\end{figure}

The primary performance metric is the aggregate downlink throughput. Figure~\ref{fig:res1} depicts the average achievable sum throughput for different network sizes. The duration of interference measurement in (\ref{dlint2}) is set as $T = n \times \mbox{Slot time}$ with $n = 1000 $. For the MPD, the probe delay is measured for the same duration and the AP with minimum probe delay is selected. From Figure~\ref{fig:res1}, as expected, we infer that under any network size, DASA improves aggregate throughput. For instance, when the number of contending STAs is 300, DASA achieved $43 \%$ (43.22 to 61.84 $\mbox{Mbps}$) and $99 \%$ (31.02 to 61.84 $\mbox{Mbps}$) throughput gains over MPD and SSF, respectively. With increasing network size, DASA outperforms existing MPD and SSF schemes with aggregate throughput approaching the optimal benchmark. Selecting AP with best DL-SINR improves the PHY rate of each AP-to-STA link, which consequently improves aggregate end-to-end throughput.

\begin{figure}[!h]
	\centering
	\includegraphics[width=5in]{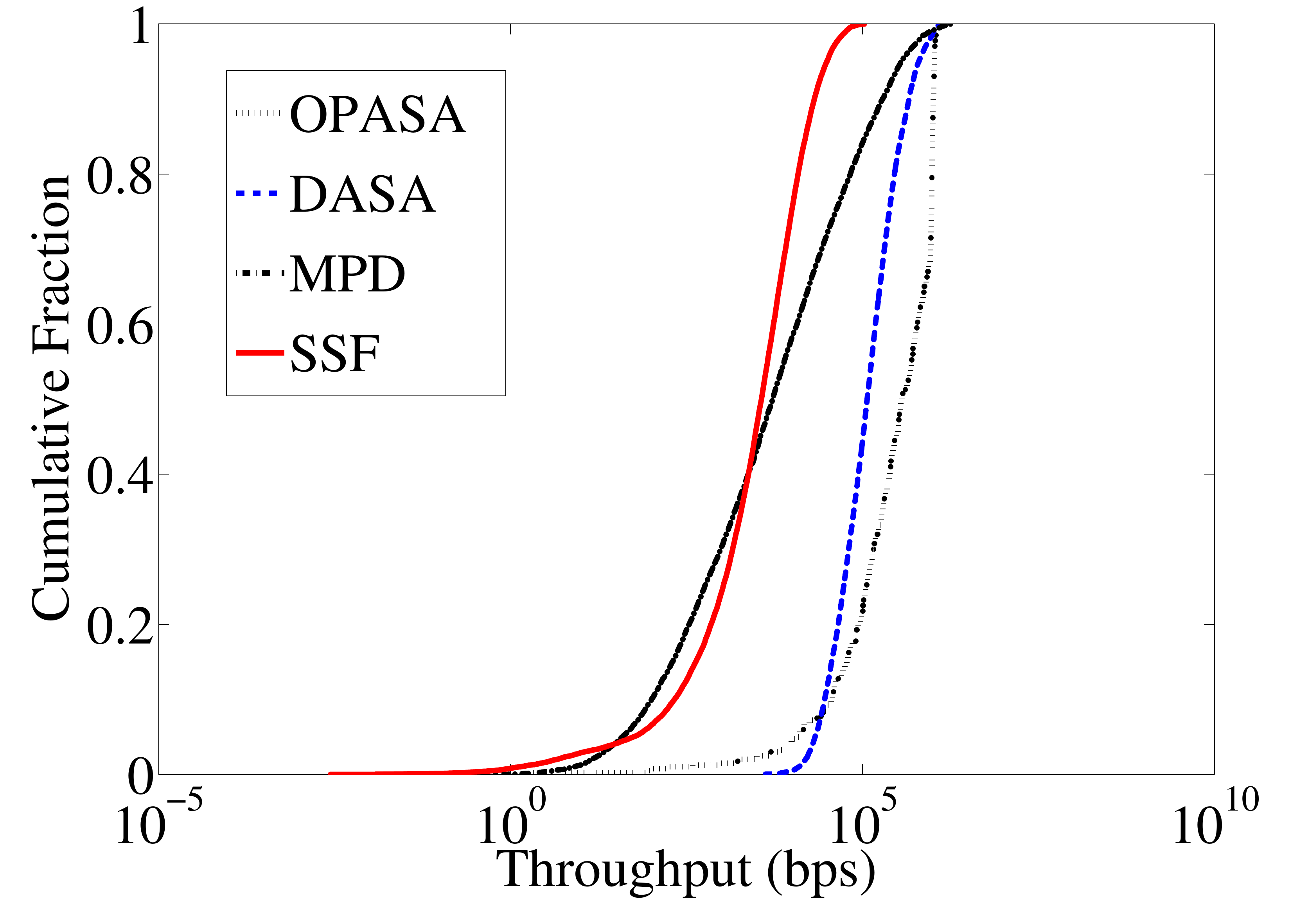}
	\caption{Per-link throughput of $400$ STAs for $n$ Slot-times, where $n = 1000$.}
	\label{fig:res2}
\end{figure}

 In Figure~\ref{fig:res2}, the cumulative distribution of all STA throughputs is presented. Between $20\mbox{th}$ and $90\mbox{th}$ percentiles, DASA obtains higher throughput closer to the optimal OPASA than MPD and SSF. Observing the 40${\mbox{th}}$ percentile, the performance of MPD over SSF fluctuates while DASA achieves nearly $2 \times $ gain over both SSF and MPD. At the $90{\mbox{th}}$ of the same Figure~\ref{fig:res2}, DASA maintains $5 \times $ gain over SSF while achieving 96.6\% gain over MPD. Between the $95{\mbox{th}}$ and $100{\mbox{th}}$ percentile throughputs under MPD and DASA schemes converge.

\begin{figure}[!h]
	\centering
	\includegraphics[width=5in]{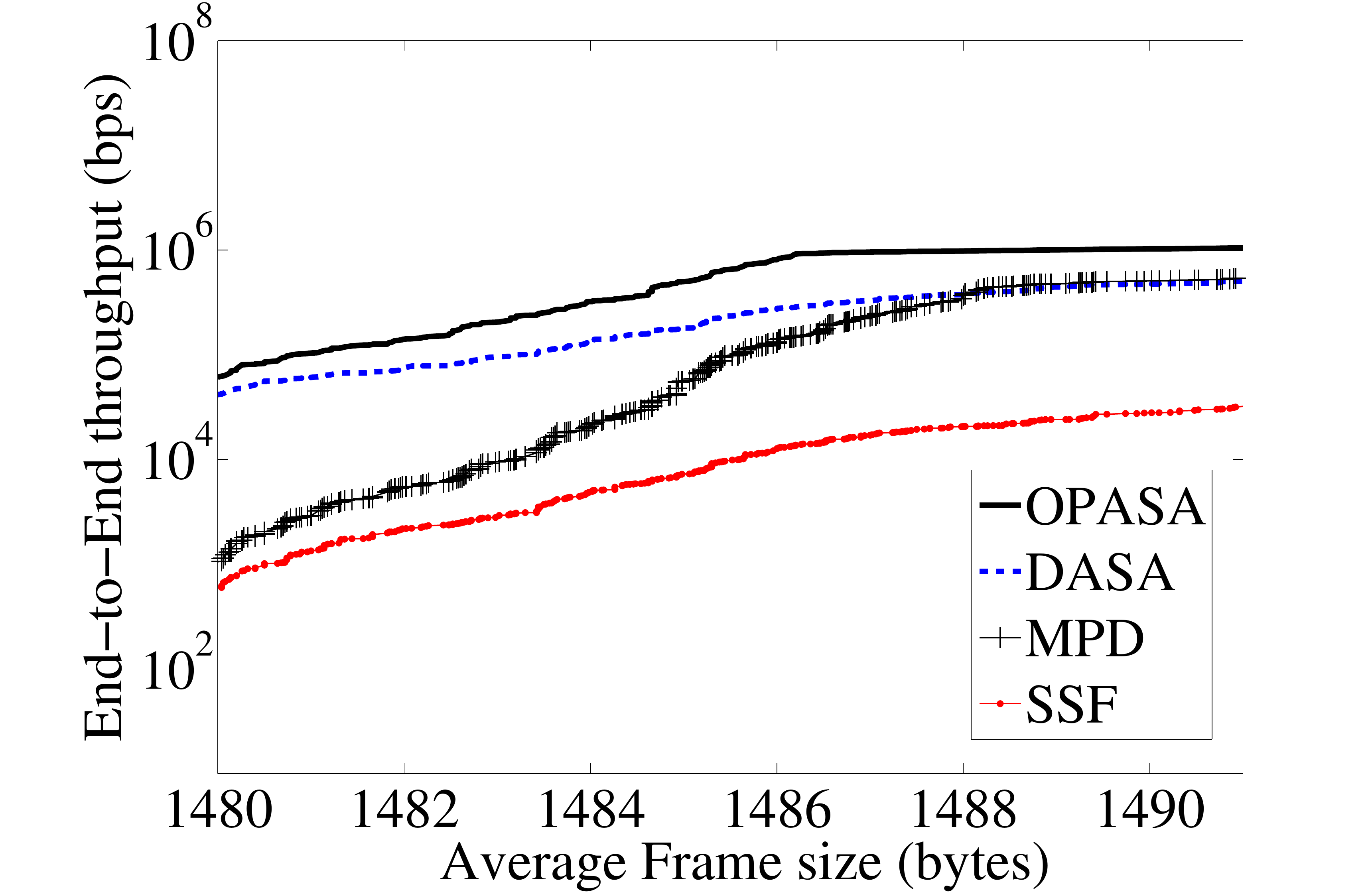}
	\caption{End-to-end throughput of $400$ STAs versus frame size.}
	\label{fig:res3}
\end{figure}

Figure~\ref{fig:res3} illustrates end-to-end throughput of each of the 400 AP-to-STA links versus frame size. The first observation in Figure~\ref{fig:res3} is that as the frame size becomes larger, the throughputs achieved under MPD and DASA converge. This is likely due to the fact that delay becomes a factor in transmitting more bits and since MPD chooses links with less delay, more bits are likely to traverse the links at the same rate in DASA. Although, both MPD and DASA significantly outperform SSF, OPASA doubles the throughputs over DASA, SSF and MPD for frame sizes below and above 1485 bytes.

\begin{figure}[!h]
	\centering
	\includegraphics[width=5in]{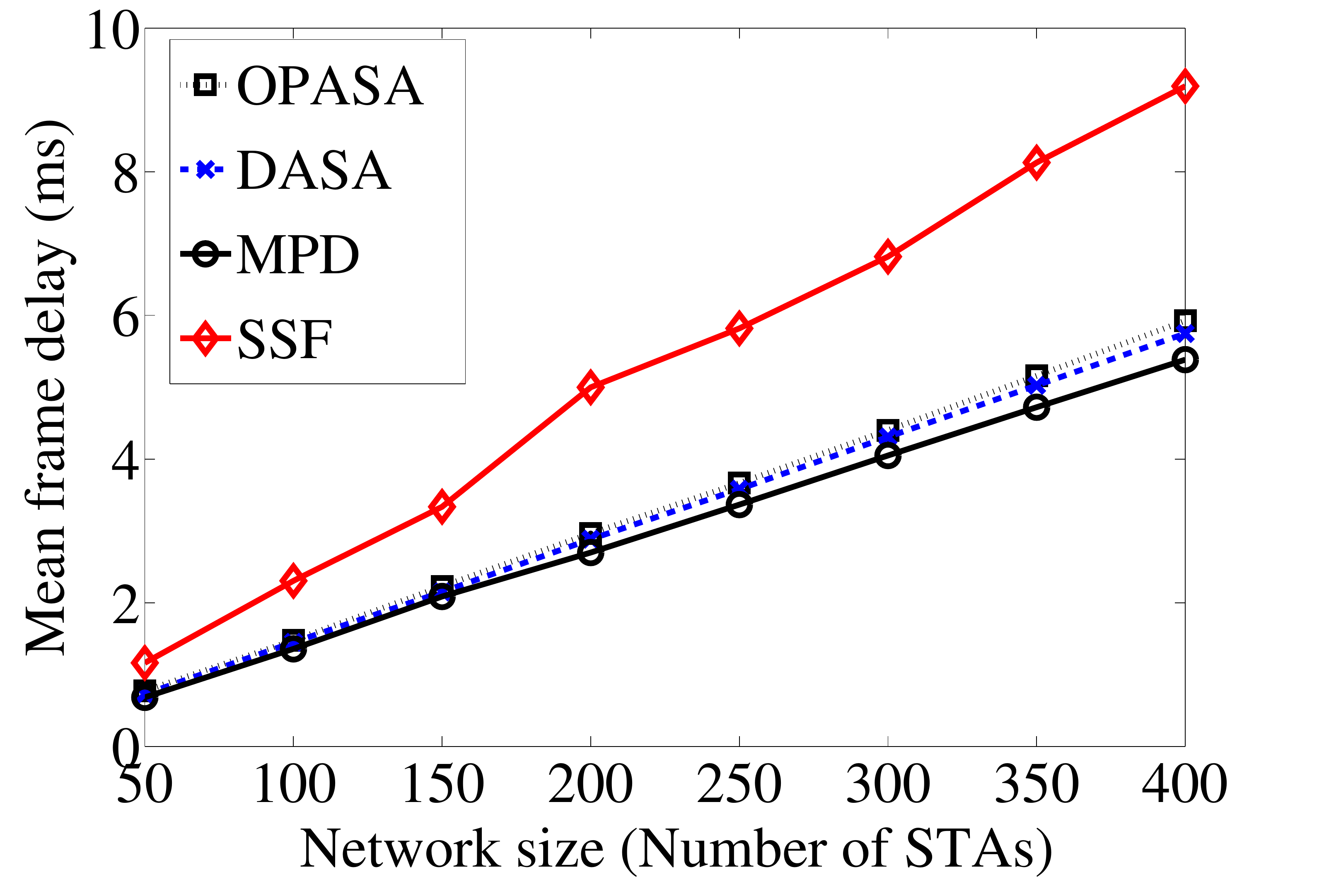}
	\caption{Mean frame delay versus network size.}
	\label{fig:res4}
\end{figure}

Figure~\ref{fig:res4} depicts the mean frame delay versus network size. Here, frame delay is the cumulative time from when a packet arrives at an AP's buffer and the AP contends for a channel, to successful reception of packet at the STA. For a small network size of 50 STAs, the delay is below $2\mbox{ms}$ for SSF, MPD, OPASA and DASA while for larger network size of 400 users, the mean delay is higher in SSF. For 400 STAs under the SSF scheme, the average frame delay is nearly $9.19\mbox{ms}$ while DASA and MPD maintain delays of $5.7\mbox{ms}$ and $5.4\mbox{ms}$, respectively. For lower network sizes (50 to 150 STAs), OPASA, MPD and DASA achieve nearly same performance in terms of delay, the slight superiority of MPD over OPASA and DASA becomes obvious when the network size increases from 200 to 400 STAs. This discrepancy is as a result of increased contentions among APs frequently trying to serve more STAs in the DL. 

%Overall, in terms of throughput, OPASA and DASA outperform SSF and MPD.

%To see the effects of each algorithm on frame delays during transmission, 

%This is the benefit of selecting an AP that offers best SINR rather than strongest RSS and less delay. 

\section{Conclusion\label{conclusion}}

The problem of inter-BSS interference is inevitable in dense 802.11 networks and degrades performance. In fact, selecting an AP with strongest RSS does not always guarantee highest throughput due to interference at the target AP. We have shown that selecting AP based on SINR reduces the effect of interference among basic service sets (BSS). This paper presents a new scheme for AP-STA  association that takes the AP interference into account. The OPASA algorithm serves as the optimal throughput benchmark while the much simpler proposed  DASA algorithm provides significant gain in aggregate throughput while taking AP interference into account. Simulation results reveal that selecting the AP offering best SINR improves throughput. Through extensive simulation, the DASA algorithm is compared to the default SSF scheme used in current 802.11 standards and the MPD algorithm proposed previously; significant throughput gain over both the SSF and the MPD schemes is observed.
 
%The problem of inter-BSS interference is inevitable in dense 802.11 networks and degrades throughput. Selecting an AP with strongest RSS does not always guarantee best throughput as interference at the target AP might be severe. This paper presents a new scheme to select an AP considering interference level in the target BSS. Our proposed algorithms OPASA and DASA provide improved end-to-end throughput because the PHY rate of each link improves when interference at a target AP is taken into account. Through extensive simulation, the proposed OPASA and DASA algorithms are compared to the default SSF scheme in current 802.11 standard and the MPD algorithm; there is significant throughput gain over both the SSF and the MPD schemes. Simulation reveals that selecting AP offering best SINR in the DL improves the DL throughput. Enhancing DL throughput is crucial as most WiFi traffic is in the DL and to reduce the effect of inter-BSS interference. 


\begin{thebibliography}{100}

\bibitem{klee} K. Lee et. al., ``Mobile data offloading: how much can WiFi deliver?,'' in {\em IEEE/ACM Trans. on Networking}, vol. 21, no. 2, April 2013.

\bibitem{shin} K. Shin, I. Park, J. Hong, D. Har and D. Cho ``Per-node throughput enhancement in Wi-Fi DenseNets,'' in {\em IEEE Comm. Magazine}, vol. 53,, no. 1, pp. 118 - 125, January, 2015.

%\bibitem{byigal} B. Yigal, H. Seung-Jae and L. Li(Erran) ``Fairness and load balancing in wireless LANs using association control''{\em IEEE/ACM Trans. on Networking}, vol. 15, no. 3, pp. 560- 573, June 2007.

\bibitem{chen} J. C. Chen et. al., ``Effective AP selection and load balancing in IEEE 802.11 wireless LANs,'' in {\em Proc. IEEE Globecom }2006.

\bibitem{weili} L. Wei et. al.,``AP association for proportional fairness in multirate WLANs,'' in {\em IEEE/ACM Trans. on Net.}, vol. 22, no. 1, Feb., 2014.

\bibitem{pbo} P. B. Oni and S. D. Blostein ``AP association optimization and CCA threshold adjustment in Dense WLANs,'' in {\em Proc. IEEE Globecom 2015 Workshop on Enabling Tech. in Future Wirel. Local Area Net.}, 2015.

\bibitem{masahiro} K. Masahiro, et al. ``A trigger-based dynamic load balancing method for WLANs using virtualized network interfaces,'' in {\em Proc. WCNC}, 2013.
%\bibitem{masahiro} K. Masahiro, T. Morihiko and Y. Keiichi ``A trigger-based dynamic load balancing method for WLANs using virtualized network interfaces,'' in {\em Proc. WCNC}, 2013.

\bibitem{leidu} D. Lei, et. al., ``QoS aware access point selection for pre-load-balancing in multi-BSSs WLAN,'' in {\em Proc. WCNC}, 2008.

\bibitem{leidu2} D. Lei, B. Yong and C. Lan ``Access point selection strategy for large-scale wireless local area networks,'' in {\em Proc. WCNC}, 2007.

\bibitem{hong} K. Hong, et al. ``Channel measurement-based access point selection in IEEE 802.11 WLANs,'' in {\em Pervasive and Mobile Computing, } 2015.

\bibitem{bjovic} Bojovic B, et. al., ``A supervised learning approach to cognitive access point selection'' {\em IEEE Globecom Workshops, 2011}.

\bibitem{xu}F. Xu, et. al., ``SmartAssoc: decentralized access point selection algorithm to improve throughput,'' in {\em IEEE Trans. on Parallel Distrib. Sys.}, vol. 24, no. 12, Dec., 2013.

%\bibitem{athan} G. Athanasiou, T. Korakis, O. Ercetin, L. Tassiulas ``A cross-layer framework for association control in wireless mesh networks,'' in {\em IEEE Trans. Mob. Comput.}, vol. 8, no. 1, 2009, pp. 65 - 80.

%\bibitem{cisco} Cisco System, Inc. ``Wireless LAN design guide for high density client environments in higher education,'' 2011, U.S.A.

\bibitem{murad} A. Murad, ``A new approach for interference measurement in 802.11 WLANs,'' {\em 21st Annual IEEE Int'l Symp. on PIMRC.}, 2010.

%\bibitem{dot11} IEEE/ANSI, ``Part 11: Wireless LAN MAC and PHY Layer Specifications,'' {\em IEEE Std 802-11-2012, 2012}.

\bibitem{gurobi} Gurobi, ``Gurobi Optimization,'' http://www.gurobi.com. Accessed: March 2nd, 2014.

%\bibitem{ref4} B. Yigal and H. Seung-Jae ``Cell breathing techniques for load balancing in wireless LANs,'' {\em IEEE Trans. on Mobile Computing}, vol. 8, no. 6, June 2009.

%\bibitem{nguyen} H. Q. Nguyen, F. Baccelli, and D. Kofman, ``A stochastic geometry analysis of dense IEEE 802.11 networks,'' in {\em Proc., IEEE INFOCOM 2007}, Anchorage , Alaska , USA.

%
%\bibitem{aibrahim} A. M. Ibrahim, T. A. ElBatt, and A. El-Keyi, ``Coverage Probability Analysis for Wireless Networks Using Repulsive Point Processes,'' CoRR, vol. abs/1309.3597, 2013.
%
%\bibitem{cisco} Cisco System, Inc. ``Wireless LAN design guide for high density client environments in higher education,'' 2011, U.S.A.
%
%\bibitem{paulo} C. Paulo, ``Modeling interference in wireless \uppercase{A}d \uppercase{H}oc networks,'' {\em IEEE Communications Surveys \& Tutorials}, vol. 12, no. 4, Fourth Quarter, 2010.
%
%\bibitem{stoyan} S. N. Chiu, D. Stoyan, W. S. Kendall, J. Mecke ``Stochastic geometry and its applications,''3rd ed. {\em John Wiley \& Sons Ltd}, 2013.


%
%\bibitem{ref2} E. G. Villegas, R. V. Ferre, and J. P. Aspas ``Load balancing in WLANs through IEEE 802.11k mechanisms'' in {\em Proc., 11th IEEE Symposium on Computers and Communications}, Pula-Cagliari, Sardinia, Italy.
%
%\bibitem{ref5} V. P. Mhatre and K. Papagiannaki ``Optimal design of high density 802.11 WLANs''{\em ACM CoNEXT}, Lisboa, Portugal, 2006.
%
%\bibitem{ref6} Z. Yanfeng, N. Zhisheng, Z. Qian, T.Bo, Z. Zhi, and Z, Jing ``A multi-AP architecture for high-density WLANs: protocol design and experimental evaluation,'' in {\em Proc., IEEE SECON 2008}, CA, U.S.A, June 16 - 20, 2008.
%
%\bibitem{ozgur} O. Ekici and A. Yongacoglu, ``A novel association algorithm for congestion relief in \uppercase{IEEE} 802.11 \uppercase{WLANs},'' in {\em Proc., ACM IWCMC`06}, Vancouver, British Columbia, Canada, July 2006.
%
%\bibitem{pbo} P. B. Oni and S. D. Blostein ``Uplink throughput enhancement in dense WLAN via association and CCA threshold optimization,'' submitted {\em Proc. IEEE GlobeCom 2015}.
%
%\bibitem{ref7} A. George, C. W. Pradeep, F. Carlo, and T. Leandros ``Optimizing client association for load balancing and fairness in millimeter-wave wireless networks,'' {\em IEEE/ACM Transactions on Networking}, 2014.
%
%\bibitem{ref8} Z. Jing, M. Benjamin, G. Xingang, and L. York ``Adaptive CSMA for scalable network capacity in high-density WLAN: a hardware prototyping approach,'' in {\em Proc., IEEE Infocom, 2006}.
%
%\bibitem{ref9} Z. Jing, G. Xingang, L. L. Yang, W. S. Conner, R. Sumit, and H. M. Mousumi ``Adapting physical carrier sensing to maximize spatial reuse in 802.11 mesh networks,'' in {\em ACM Journal, Wireless Communications \& Mobile Computing - Special Issue: Emerging WLAN Applications and Technologies archive}, vol. 4 Issue 8, December 2004, pp. 933 - 946
%
%\bibitem{ref10} R. E. Burkard and E. Cela ``Handbook of combinatorial optimization, supplement volume A''{\em Kluwer Academic Publishers}, 1999, ch. Linear assignment problems and extensions, pp. 75 - 149.
%
%\bibitem{ref11} J. C. Setubal ``Sequential and parallel experimental results with bipartite matching algorithms''{\em Technical Report IC-96-09}, Institute of Computing, State University of Campinas, Brazil, 1996.
%
%\bibitem{ref14} N. J. A. Harvey, R. E. Ladner, L. Lovasz, and T. Tamir, ``Semi-matchings for bipartite graphs and load balancing'' in {\em Journal of Algorithms}, Academic Press, vol. 59, Issue 1, April 2006, pp. 53 - 78
%
%\bibitem{ref15} V. L. Hai and S. Taka, ``Collision probability in saturated IEEE 802.11 networks,'' in {\em Australian Telecommunication Networks and Applications Conference, 2006}.
%
%\bibitem{ref16} K. Ponnusamy and A. Krishnan, ``Saturation throughput analysis of IEEE 802.11b wireless local area networks under high interference considering capture effects,'' {\em International Journal of Computer Science and Information Security, vol. 7, no. 1, 2010}.
%
%\bibitem{ref17} A. Aditya, J. Glenn, S. Srinivasan, and S. Peter, ``Self-management in chaotic wireless deployments,'' in {\em MobiCom'05, Germany, 2005}.
%
%\bibitem{ref18} IEEE/ANSI, ``Wireless LAN MAC and PHY Layer Specifications,'' {\em IEEE Std 802-11-1999, 1999}.
%
%\bibitem{ref19} B. O'Hara, ``IEEE 802.11 handbook: a designer companion,'' {\em IEEE Standards Information Network, IEEE Press, NY, USA, 2005}.
%
%\bibitem{ref20} B. Akash, S. Michael, S. Ivan and R. Jennifer and R. Dipankar, ``Network cooperation for client-\uppercase{AP} association optimization,'' in {\em Proc., 10th International Symp. on Modeling and Optimization in Mobile, \uppercase{A}d Hoc and Wireless Networks (WiOpt)}, Germany, 2012.
%
%\bibitem{ref26} O. B. Karimi, J. Liu, and J. Rexford ``Optimal collaborative access point association in wireless networks,'' in {\em Proc.}, IEEE INFOCOM'14, Toronto, April 27-May 2, 2014.
%
%\bibitem{murad} M. Abusubaih and A. Wolisz ``An optimal station association policy for multi-rate \uppercase{IEEE} 802.11 wireless \uppercase{LAN}s,'' in {\em Proc., ACM MSWiM`07}, Crete Island, Greece, October, 2007.
%
%\bibitem{aditya} G. Aditya and K. Sachin ``Strider: automatic rate adaptation and collision handling,'' in {\em Proc., IEEE SIGCOMM'11}, Toronto, Canada, August, 2011.
%
%\bibitem{kumar} A. Kumar and V. Kumar ``Optimal association of stations and APs in an \uppercase{IEEE} 802.11 \uppercase{WLAN},'' in {\em Proc., National Communications Conference (NCC)}, Bangalore, India, January, 2005.
%
%\bibitem{ioannis} I. Koukoutsidis and V. A. Siris ``Access point assignment algorithms in \uppercase{WLANs} based on throughput objectives,'' in {\em Proc., IEEE Symp. on Modeling and Optimization in Mobile, Ad Hoc, and Wireless Networks and Workshops (WiOPT 2008)}, Berlin, Germany, April, 2008.
%
%\bibitem{gong} H. Gong, K. Nahm and J. W. Kim ``Access point selection tradeoff for \uppercase{IEEE} 802.11 wireless mesh network,'' in {\em Proc., IEEE Consumer Communications and Networking Conference (CCNC 2007)}, Las Vegas, NV, USA, January, 2007.
%
%\bibitem{yen} L. Yen, J. Li and C. Lin ``Stability and fairness of \uppercase{AP} selection games in \uppercase{IEEE} 802.11 access networks,'' in {\em IEEE Trans. on Vehicular Tech.}, vol. 60,, no. 3, pp. 1150 - 1160, March, 2011.
%
%\bibitem{lili} L. Li, M. Pal and Y. R. Yang ``Proportional fairness in multi-rate wireless \uppercase{LAN}s,'' in {\em Proc., IEEE INFOCOM`08}, Phoenix, USA, March, 2008.
%
%\bibitem{zhu} J. Zhu, X. Guo, S. Roy and K. Papagiannaki ``\uppercase{CSMA} self-adaptation based on interference differentiation,'' in {\em Proc., IEEE GLOBECOM`07}, Washington, DC, USA, November, 2007.
%
%\bibitem{imad} I. Jamil, L. Cariou and J. H\'{e}lard ``Improving the capacity of future \uppercase{IEEE} 802.11 high efficiency \uppercase{WLAN}s,'' in {\em Proc., IEEE Intl. Conf. on Telecom.}, Lisbon, Portugal, May, 2014.
%
%\bibitem{jzhu} J. Zhu, X. Guo, L. L. Yang, W. S. Conner, S. Roy and M. M. Hazra ``Adapting physical carrier sensing to maximize spatial reuse in 802.11 mesh networks,'' in {\em Wireless Communications and Mobile Computing - Special Issue: Emerging WLAN Apllications and Technologies archive}, vol. 4,, no. 8, pp. 933 - 946, December, 2004.
%
%\bibitem{jason} J A. Fuemmeler , N. H. Vaidya and V. V. Veeravalli ``Selecting transmit powers and carrier sense thresholds in \uppercase{CSMA} protocols for wireless \uppercase{A}d \uppercase{H}oc networks,'' in {\em Proc., ACM Intl. Wireless Internet Conf.(WICON`06)}, Boston, MA, USA, August, 2006.
%
%\bibitem{zzhou} Z. Zhou, Y. Zhu, Z. Niu and J. Zhu ``Joint tuning of physical carrier sensing, power and rate in  \uppercase{H}igh-\uppercase{D}ensity \uppercase{WLAN},'' in {\em Proc., IEEE Asia-Pacific Conf. on Communications}, Bangkok, October, 2007.
%
%\bibitem{shin} K. Shin, I. Park, J. Hong, D. Har and D. Cho ``Per-node throughput enhancement in Wi-Fi DenseNets,'' in {\em IEEE Comm. Magazine}, vol. 53,, no. 1, pp. 118 - 125, January, 2015.
%
%\bibitem{waypoint} C. Bettstetter, H. Hartenstein and X. Perez-Costa ``Stochastic properties of the random waypoint mobility model,'' in {\em Wireless Networks}, vol. 10,, no. 5, pp. 555-657, 2004.
%
%\bibitem{beatriz} B. Soret, K. I. Pedersen, N. T. K J{\o}rgensen and V. Fern\'{a}dez-L\'{o}pez ``Interference coordination for dense wireless networks,'' in {\em IEEE Comm. Magazine}, vol. 53,, no. 1, pp. 102 - 109, January, 2015.
%
%\bibitem{weili2} L. Wei, C. Yong, C. Xuizhen, A. A. Mznah and A. Abdullah ``Achieving proportional fairness via \uppercase{AP} power control in multi-rate \uppercase{WLANs},'' in {\em IEEE Trans. on Wireless Comm.}, vol. 10,, no. 11, pp. 3784 - 3792, February, 2011.
%
%\bibitem{siris} I. Koukoutsidis and V. A. Siris, ``Access point assignment algorithms in WLANs based on throughput objectives,'' {\em IEEE Symp. WiOPT}, 2008, 1 - 3. April, 2008, Berlin, Germany.
%
%\bibitem{ref27} ITU-R, Recommendation ITU-R P.1238-4 ``Propagation data and prediction methods for the planning of indoor radiocommunication systems and radio local area networks in the frequency range 900 MHz to 100 GHz,'' 2012.
%
%\bibitem{ismail} I. H. Toroslu and G. \"{U}\c{c}çoluk ``Incremental assignment problem,'' in {\em Journal of Information Sciences}, vol. 117, no. 6, 2007, pp. 1523 – 1529
%
%\bibitem{weili} L. Wei, W. Shengling, C. Yong, C. Xuizhen, X. Ran, A. A. Mznah, and A. Abdullah ``AP association for proportional fairness in multirate WLANs,'' in {\em IEEE/ACM Trans. on Networking}, vol. 22, no. 1, February, 2014.
%
%\bibitem{mills} G. A. Mills-Tettey, A. Stentz, and M. B. Dias, ``The dynamic Hungarian algorithm for the assignment problem with changing costs,'' Robotics Institute, School of Computer Science, Carnegie Mellon University, PA, July 2007.
%
%\bibitem{lawler} E. L., Lawler, ``Combinatorial Optimization: Networks and Matroids,'' Holt, Reinehart and Winston, New York, USA, 1976.
%
%\bibitem{steiglitz} C. H., Papadimitriou, and K., Steiglitz, ``Combinatorial Optimization: Algorithms and Complexity,'' Dover Publications, U.S.A, 1998.



\end{thebibliography}
\end{document}